\documentstyle[11pt,aaspp4]{article}

\def\be{\begin{equation}}
\def\ee{\end{equation}}
\def\eqref#1{(\ref{eqn:#1})}

\def\hmpc{{h^{-1}\;{\rm Mpc}}}

\def\kms{{\rm \;km\;s^{-1}}}

\def\lya{Ly$\alpha$ }
\def\lyb{Ly$\beta$ }
\def\lyg{Ly$\gamma$ }
\def\lyd{Ly$\delta$ }
\begin{document}

\slugcomment{to be published in the Astronomical Journal}

\title{Evolution of the Ionizing Background and the Epoch of Reionization from the Spectra of $z\sim 6$ Quasars}

\author{Xiaohui Fan\altaffilmark{\ref{IAS}},
Vijay K. Narayanan\altaffilmark{\ref{Princeton}},
Michael A. Strauss\altaffilmark{\ref{Princeton}},
Richard L. White\altaffilmark{\ref{STScI}},
Robert H. Becker\altaffilmark{\ref{UCDavis},\ref{IGPP}}
Laura Pentericci\altaffilmark{\ref{Heidelberg}},
Hans-Walter Rix\altaffilmark{\ref{Heidelberg}}
}
\newcounter{address}
\setcounter{address}{1}
\altaffiltext{\theaddress}{Institute for Advanced Study, Olden Lane,
Princeton, NJ 08540
\label{IAS}}
\addtocounter{address}{1}
\altaffiltext{\theaddress}{Princeton University Observatory, Princeton,
NJ 08544
\label{Princeton}}
\addtocounter{address}{1}
\altaffiltext{\theaddress}{Space Telescope Science Institute, Baltimore, MD 21218
\label{STScI}}
\addtocounter{address}{1}
\altaffiltext{\theaddress}{Physics Department, University of California, Davis,
CA 95616
\label{UCDavis}}
\addtocounter{address}{1}
\altaffiltext{\theaddress}{IGPP/Lawrence Livermore National Laboratory, Livermore, CA 94550
\label{IGPP}}
\addtocounter{address}{1}
\altaffiltext{\theaddress}{Max-Planck-Institut f\"{u}r Astronomie,
K\"{o}nigstuhl 17, D-69171 Heidelberg, Germany
\label{Heidelberg}}

\begin{abstract}
We study the process of cosmic reionization and
estimate the ionizing background in the intergalactic medium (IGM)
using the Lyman series absorption in the spectra of the four 
quasars at $5.7 < z < 6.3$ discovered by the Sloan Digital Sky Survey.
We derive the redshift evolution of the ionizing background at 
high redshifts, using both semi-analytic techniques and cosmological 
simulations to model the density fluctuations in the IGM.
The existence of the complete \lya Gunn-Peterson trough
in the spectrum of the $z=6.28$ quasar SDSS 1030+0524
indicates a photoionization rate ($\Gamma_{-12}$ in units of
$10^{-12} {\rm s}^{-1}$) at $z \sim 6$ lower than $0.08$,  at least
a factor of $6$ smaller than the value at $z\sim 3$.
The \lyb and \lyg Gunn-Peterson troughs give an even
stronger limit $\Gamma_{-12} \lesssim 0.02$ due to their smaller
oscillator strengths, indicating that the ionizing background in the
IGM at $z\sim 6$ is more than 20 times lower than that at $z\sim 3$.
Meanwhile, the volume-averaged neutral hydrogen fraction increases
from $10^{-5}$ at $z\sim 3$ to $>10^{-3}$ at $z\sim 6$.
At this redshift, the mass-averaged neutral hydrogen fraction is larger than 1\%;
the mildly overdense regions ($\delta > 3$) 
are still mostly neutral and the comoving mean free path of ionizing 
photons is shorter than 8 Mpc.
Comparison with simulations of cosmological reionization 
shows that the observed properties
of the IGM at $z\sim 6$ are typical of those in the era at the end of
the overlap stage of reionization when the individual HII regions merge.
Thus, $z\sim 6$ marks the end of the reionization epoch. 
The redshift of reionization constrains the small scale power of the 
mass density fluctuations and the star forming
efficiency of the first generation of objects. 

\end{abstract}

\section{Introduction}
After the recombination epoch at $z\sim 1500$, the universe remained
mostly neutral, until the first generation of stars and quasars
ionized the intergalactic medium (IGM) and ended the 
cosmic ``dark ages'' (e.g. \cite{Rees98}).
Popular cosmological models predict this reionization to have occurred
at redshift between 6 and 20 (e.g., \cite{GO97}, Chiu \& Ostriker 2000, 
\cite{Nick00}, \cite{Ciardi01}, \cite{Razoumov01} and references therein).
When and how the universe reionized is one of the fundamental
questions of modern cosmology (for a complete review, see \cite{BL01} 
and \cite{LB01}). 
Observations of absorption spectra of luminous sources at high 
redshifts approaching the reionization epoch provide the best 
observational probes of reionization to date. 

Recent discoveries of luminous quasars at $z >  5.7$
(\cite{Fan00}, \cite{PaperI}) using imaging data from the
Sloan Digital Sky Survey (SDSS, \cite{York00}) enable  us to study the 
state of the intergalactic medium (IGM) at redshift up to 6.
Observations of Fan et al. (2001c, hereafter Paper I), 
Becker et al. (2001, hereafter Paper II) 
and Pentericci et al.  (2001, hereafter Paper III) 
show that the \lya absorption due to neutral hydrogen in the IGM
increases dramatically toward high redshifts.
In particular, Keck  
and VLT 
spectroscopy of high redshift quasars
(in Papers II and III, respectively) show the first observation of a complete 
Gunn-Peterson (GP) trough (Shklovsky 1964, Scheuer 1965, 
Gunn \& Peterson 1965) in the spectrum of the $z=6.28$ quasar, 
SDSSp J103027.10+052455.0 (SDSS 1030+0524 for brevity),
where no flux is detected over a 300\AA\ region immediately blueward
of the \lya emission line.
The flux decrement between the red and blue sides of the 
\lya emission line is larger than $150$, 
corresponding to an effective optical depth to \lya photons
($\tau_{\rm eff}$) larger than 5 at $z_{\rm abs} \sim 6.05$.
The existence of a \lyb GP trough in the Keck
spectrum imposes an even stronger limit on the effective equivalent
\lya optical depth ($\tau_{\rm eff} > 20$, Paper II).
\cite{D01} also observed a particularly dark region of length 
$\sim 5$ Mpc at $z \sim 5.4$, along the line of sight to the 
$z\sim 5.8$ quasar SDSS 1044--0125.

This is a theoretical companion paper to Papers I -- III. In this paper,
we use the absorption measurements presented in the previous
papers to calculate the evolution of the ionizing background and the
neutral hydrogen fraction in the IGM, and constrain the epoch of
reionization. The outline of this paper is as follows.
In \S2, we summarize the observed evolution of the average \lya
absorption along the lines of the sight to the sample of quasars in Paper I
based on the spectra presented in Papers II and III,
and discuss the errors on these measurements. 
In \S3, we calculate the ionizing background at different redshifts using
the average \lya absorption, taking into account the inhomogeneity of 
the IGM using a model for the overdensity distribution.
We derive stronger constraints on the ionizing background in the
GP trough region of SDSS 1030+0524 at $z_{\rm abs} \sim 6$ using the
\lyb and \lyg absorption troughs.
In \S4, we use N-body simulations to model the high-redshift
\lya forest spectra, and compare the statistical properties
of the simulated and observed spectra.
In \S5, we calculate the redshift evolution of the volume and mass
weighted neutral hydrogen fractions in the IGM, and show how the 
reionization penetrates  into progressively
overdense regions at lower redshifts as the mean free path of ionizing photons
increases.
These calculations are used to constrain the epoch of reionization.
We discuss the prediction of  
structure formation models for the reionization epoch, 
and outline future observations needed to probe deep into and beyond
the reionization epoch in \S6.

Throughout the paper, we adopt a LCDM cosmology with $\Omega_m = 0.35$, 
$\Lambda = 0.65$, $h = 0.65$, and $\Omega_b h^2 = 0.02$
unless otherwise noted.

\section{Evolution of Neutral Hydrogen Absorption}

The Gunn-Peterson (1965) optical depth to \lya photons is
\begin{equation}
\tau_{\rm GP} = \frac{\pi e^2}{m_e c} f_{\alpha} \lambda_{\alpha} H^{-1}(z) n_{\rm HI },
\end{equation}
where $f_{\alpha}$ is the oscillator strength of the \lya transition,
$\lambda_\alpha$ = 1216\AA, $H(z)$ is the Hubble constant at redshift
$z$, and $n_{\rm HI }$ is the density of neutral hydrogen in the IGM.
At high redshifts, $H(z) \propto  h\Omega_m^{1/2}(1+z)^{3/2}$, and the GP
optical depth for a uniformly distributed IGM can be re-written as:
\begin{equation}
\tau_{\rm GP} (z) = 1.8 \times 10^5 h^{-1} \Omega_m^{-1/2}
\left( \frac{\Omega_b h^2}{0.02} \right)
\left ( \frac{1+z}{7} \right )^{3/2}
\left( \frac{n_{\rm HI}}{n_{\rm H}} \right ).
\end{equation}
In reality, the IGM is highly inhomogeneous, and regions with different 
densities will be in  different ionization states and 
have different GP optical depths (\S3). 
The IGM density distribution needs to be taken into account when estimating
the averaged neutral fraction from the observations (\S5).

Figure 1 shows the observed evolution of the average \lya absorption
in the high-redshift quasar spectra as a function of redshift, 
both in terms of the transmitted flux ratio $\cal{T}$, 
defined  as the ratio of observed and unabsorbed continuum fluxes 
(see Paper I),
\begin{equation}
{\cal T}(z_{\rm abs}) \equiv \left\langle f_\nu^{\rm obs}/f_\nu^{\rm con} \right\rangle,
\end{equation}
and the effective GP optical depth,  defined as  
\begin{equation}
\tau_{\rm GP}^{\rm eff}  \equiv - \ln({\cal T}).
\end{equation}

The measurements shown in Figure 1 at $z_{\rm abs}<4.5$  are taken from 
\cite{MM01}. 
The measurements at $4.8 < z_{\rm abs} <  5.6$ are
based on the spectra of Paper II and III. 
In this Figure, we have averaged the measurements along four lines of
sights into four different redshift bins, 
$z_{\rm abs}=$ [4.8 -- 5.2], [5.2 -- 5.6], 
[5.6 -- 5.95] and [5.95 -- 6.15], respectively, 
and include the systematic errors from continuum shape and sample
variance (see below).
We compute these quantities using only the data between 
$z_{\rm abs} < z_{\rm em} - 0.1$ and 
$ 1 + z_{\rm abs} > 1040 \times ( 1+z_{\rm em})/1216$, to avoid 
contamination from the quasar \lya emission line and the proximity effect on
the red side, and from the quasar \lyb+OVI emission lines on the blue side.
The value in the highest redshift bin, $z = $ [5.95 -- 6.15], 
${\cal T} = 0.004 \pm 0.003$, is consistent with
zero transmitted flux. 
Figure 1(b) shows the 1-$\sigma$ {\em lower} limit on the
optical depth $\tau_{\rm GP}^{\rm eff} > 5.1$ from the \lya measurement.

The error on the transmitted flux ratio
includes contributions from three sources: (1) photon noise in
the spectrum, $\sigma(\rm photon)$, which is measured directly
from the data. 
The typical $\sigma(\rm photon)$ is of the order $2 \times 10^{-3}$. 
(2) systematic uncertainties due to the extrapolation
of the intrinsic quasar continuum, $\sigma(\rm con)$.
In Paper I--III, we assumed a power law continuum 
$f_\nu \propto \nu^{-\alpha}$,
where $\alpha = 0.5$, normalized at rest frame wavelength 1280\AA. 
Therefore,
\begin{equation}
\sigma_{\rm con}  = \left| \log \left[\frac{1216(1+z_{\rm abs})}{1280(1+z_{\rm em})} \right] \right|
 \sigma(\alpha) \times {\cal T}.  
\end{equation}
For $\sigma(\alpha) \sim 0.4$ (Fan et al. 2001a), 
the typical relative error is $\sigma_{\rm con}/{\cal T} \sim 0.05 $. 
(3) sample variance, $\sigma(\rm sample)$.
This quantity can be roughly estimated using a model in which the
\lya clouds are distributed randomly in the IGM (Zuo 1992).
We calculate $\sigma(\rm sample)$ using
Equations (8) and (9) of Zuo (1992), assuming that
the column density distribution of the \lya clouds follows a power law 
$N(N_{HI}) \propto N_{HI}^{-\beta}$, with $\beta = -1.5$, and the 
Doppler width of the \lya forest line is $b = 30$ km s$^{-1}$.
In a redshift window $\delta z = 0.4$, 
$\sigma(\rm sample) \sim 0.025$ for ${\cal T} \sim 0.10$ while
for $\delta z = 0.2$, $\sigma(\rm sample) \sim 0.002$ for 
${\cal T} \sim 0.006.$
Therefore, for all the measurements except at the highest redshift 
$(z_{\rm abs} \sim 6$),
the error term is dominated by sample variance. This conclusion is
consistent with the scatter of ${\cal T}$ measurements from
different lines of sight at the same redshift, which is
larger than that expected from photon noise alone.
For ${\cal T} \ll 1$, the contribution from photon noise becomes important.
The error bars in Figure 1 represent the sum of the three error terms,
added in quadrature.

\section{Evolution of the Ionizing Background}

The \lya forest arises from absorption of \lya photons by neutral hydrogen gas
in the intergalactic medium (eq. 2). Assuming local photoionization-recombination 
equilibrium, we have:
\begin{equation}
n_{\rm HI} \Gamma = n_{\rm HII} n_e \alpha(T),  
\end{equation} 
where $n_{\rm HI}$, $n_{\rm HII}$ and $n_e$ are the {\em local} densities of
neutral,  ionized hydrogen and 
electrons in the IGM, respectively, 
$\Gamma$ is the photoionization rate, and $\alpha(T)$ is the
recombination coefficient at temperature $T$ (Abel et~al. 1997),
\begin{equation}
\alpha(T) = 4.2\times10^{-13} (T/10^4\rm K)^{-0.7}
{\rm cm}^{3} {\rm s}^{-1}.
\end{equation}
The photoionization rate $\Gamma$ is related to the ionizing
background flux  $J_{\nu}$ by,
\begin{equation}
\Gamma = 4 \pi \int \frac{J_\nu}{h\nu} \sigma_\nu d\nu,
\end{equation}
where the integral includes the ionizing photons from the HI Lyman Limit
to the HeII Lyman Limit assuming that the He in the IGM is singly ionized, and
$\sigma_\nu$ is the HI cross section of ionizing photons,
$\sigma_\nu \sim \nu^{-3}$.
For a power law spectrum of the ionizing background dominated by quasar,
$J_\nu \propto \nu^{-1.8}$ (\cite{MHR99}),
we find $\Gamma_{-12} = 2.5 J_{-21}$,
where $\Gamma_{-12}$ is the photoionization rate in units of 
$10^{-12}$ s$^{-1}$, and
$J_{-21}$ is the ionizing flux at the Lyman Limit in units of 
$10^{-21}$ erg cm$^{-2}$ s$^{-1}$ Hz$^{-1}$ sr$^{-1}$.
If the ionizing background at high-redshifts is dominated by massive stars 
(e.g. Fan et al. 2001c) with 
$J_\nu \propto \nu^{-5}$ (\cite{BL01}),
we find $\Gamma_{-12} = 1.5 J_{-21}$.

If the IGM is mostly ionized by a uniform ionizing background,
the evolution of the optical depth can be expressed as (Weinberg et al. 1997),
\begin{equation}
\tau_{\rm GP} \propto \frac{(1+z)^6 (\Omega_b h^2)^2 \alpha(T)}{\Gamma H(z)}
\propto \frac{(1+z)^{4.5} (\Omega_b h^2)^2 \alpha(T)}{h \Gamma \Omega_m^{0.5}}.
\end{equation}
The \lya absorption increases 
rapidly with increasing redshift 
even if the ionizing background remains constant with redshift.
\cite{MM01} estimate the ionizing background  at $z<5.2$ by comparing
the observed transmitted flux ratio to that of artificial \lya forest
spectra created from cosmological simulations.
In this section, 
we develop a semi-analytic model which uses an empirical
model for the density distribution derived from 
hydrodynamical simulations similar to that of \cite{MM01}, and use it to
estimate the ionizing background from the \lya
observations (\S3.1).
Then, we use the \lyb and \lyg forest
GP trough measurements in the spectrum of SDSS 1030+0524 
to put stronger constraints on the background at $z\sim 6$ (\S3.2). 

\subsection{Semi-analytic Modelling of \lya Absorption}

The \lya forest arises from low-density 
gas in the IGM that is in approximate thermal equilibrium
between photoionization heating by the UV background and adiabatic 
cooling due to Hubble expansion  
(see e.g., \cite{bi93}; \cite{cen94}; \cite{zhang95}; \cite{hernquist96}; 
\cite{hui97}).
The neutral hydrogen fraction and therefore the GP optical depth 
depends on the local density of the IGM.
We define the fractional density of the IGM as 
$\Delta \equiv \rho/\langle \rho \rangle$ ($\equiv \delta + 1$, where 
$\delta$, the density contrast, is the departure of the local density
from the mean density, in units of the mean density).
For a region of IGM with density $\Delta$
\begin{equation}
\tau(\Delta) \propto \frac{(1+z)^{4.5} (\Omega_b h^2)^2 \alpha[T(\Delta)]}{ h\Gamma(\Delta,z) \Omega_m^{0.5}} \Delta^2. 
\end{equation}
The dependence on $\Delta^2$ arises because $\tau \propto n_{\rm HI}$, which
is proportional to $n_{\rm HII}^2$ (eq.6), and proportional to $\Delta^2$ for a highly ionized
IGM. 
The temperature of the IGM is determined by 
photoionization-recombination equilibrium, which leads 
to a power-law relation between temperature and density, of the form 
$T = T_0 \Delta^{\gamma}$, with $T_{0} \sim 1 - 2 \times
10^4$ K, and $\gamma \sim 0 - 1$ (e.g. \cite{hui97}).
Following \cite{MM01}, we assume an uniform ionizing background, and
$\gamma = 0$. Thus,
\begin{equation}
\tau(\Delta) = \tau_0 \left( \frac{1+z}{7} \right)^{4.5} \left(
\frac{0.05}{\Gamma_{-12}(z)} \right) \Delta^2.
\end{equation}
We determine $\tau_{0}$ below.
The mean transmitted flux ratio ${\cal T}$ can be calculated as
\begin{equation}
{\cal T} = \langle e^{-\tau} \rangle 
= \int_0^\infty e^{-\tau(\Delta)} p(\Delta) d(\Delta),
\end{equation}
where $p(\Delta)$ is the distribution function of the density of the
IGM.
For an inhomogeneous IGM, using the definition of effective 
optical depth in Eq.(4) and (12),
we have ${\cal T} = e^{-\tau_{\rm eff}} = \langle e^{-\tau} \rangle > e^{-\langle \tau \rangle}$,
thus $\tau_{\rm eff} < \langle \tau \rangle$.

We calculate ${\cal T}$ using a parametric form for 
the volume-weighted density distribution function $p(\Delta)$  
(Miralda-Escud\'{e}, Haehnelt, \& Rees, 2000):
\begin{equation}
p(\Delta) = A \exp \left[ - \frac{(\Delta^{-2/3} - C_0)^2}
		{2(2\delta_0/3)^2}\right] \Delta^{-\beta},
\end{equation}
where $\delta_0 = 7.61/(1+z)$, and $\beta$, $C_0$ and $A$ are numerical
constants given in Table 1 of Miralda-Escud\'e et al. (2000)  
at several redshifts.
This distribution is derived assuming that the
initial density fluctuations form a Gaussian random field, the gas in voids 
is expanding at a constant velocity, and the density field is smoothed 
on the Jeans scale of the photoionized gas.
It reproduces well the density distribution of the photoionized gas
in the LCDM hydrodynamic simulations of Miralda-Escud\'e et al. (1996).
In the following calculations, we linearly interpolate $\beta$
in redshift, and set the constants $C_0$ and $A$ so that
both the total volume and mass are normalized to unity.
Finally, we find the numerical constant in Eq.(11) to be 
$\tau_0 = 82$ by requiring the resulting $\Gamma$ 
to match the measurement of \cite{MM01} at $z=4.5$.
Note that the quantity we
are really constraining with \lya absorption is the combination
$\Gamma \Omega_{m}^{1/2}\Omega_b^{-2} h^{-3}T_0^{0.7}$ (equations 10 and 11).
There are still considerable uncertainties in the values of
$\Omega_{m}$, $\Omega_b$, $h$ and $T_0$, all of which translate to
uncertainties in the normalization of $\Gamma_{12}$.
However, in the era after reionization, the temperature of the IGM
$(T_0)$  is not a strong function of redshift.
Therefore, the redshift {\em evolution rate} of $\Gamma_{12}$ derived here
is robust to uncertainties in the normalization.
Note that \cite{MM01} assume that $T_0 = 2\times 10^4$K,
$\Omega_{b}h^{2} = 0.02$, $h=0.65$ and $\Omega_{m} = 0.4$.
The $\Gamma_{12}(z)$ would be $\sim 40$\% higher for $T_0 = 10^4$K
(e.g., \cite{Schaye00}).
\cite{MM01} show that the estimate of $\Gamma_{12}$ is quite
insensitive to the slope of the $\Delta - T$ relation
(i.e., the value of $\gamma$).

Also note that the effective optical depth here is calculated by
integrating in real space rather than in redshift space.
We compared the results from N-body simulations
using redshift and real space coordinates (\S4) 
and found that the effect of ignoring peculiar velocities 
on the resulting optical depth is negligible.

Figure 2 shows the evolution of the photoionization rate $\Gamma_{-12}$
based on the observations presented in Figure 1, including
the measurements at $z<4.5$ from \cite{MM01}.
The ionizing background declines toward higher redshift,
from $\Gamma_{-12} \sim 0.5$ at $z_{\rm abs}\sim 3$,
to $\Gamma_{-12} \sim 0.2$ at $z_{\rm abs}\sim 5$,
and to $\Gamma_{-12} \sim 0.12$ at $z_{\rm abs}\sim 5.8$.
Note that $\Gamma$ remains roughly constant between 
$z\sim 4.4 - 5.5$, albeit with rather large error bars
on individual measurements (see also \cite{CM01}).
In the GP trough region in the highest redshift object,
SDSS 1030+0524, the upper limit on the transmitted flux ratio,
${\cal T} < 0.006$, yields a much more stringent {\em upper} limit on
the ionizing background at $z_{\rm abs} \sim 6.05$: $\Gamma_{-12} < 0.08$,
a factor of 6 smaller than the measured value at $z_{\rm abs} \sim 3$.
This dramatic decline in the ionizing background indicates a much
larger neutral fraction of the IGM and a much shorter mean free path
of the ionizing photons at $z_{\rm abs} = 6$ that at $z_{\rm abs} = 3$ (\S 5).

\subsection{Constraints from \lyb and \lyg Troughs in SDSS 1030+0524}

As pointed out in Paper II, because of its much smaller oscillator strength,
and the fact that the flux scales exponentially with oscillator strength, 
one can put a stronger limit on the neutral hydrogen fraction in the 
GP trough from the \lyb transition than from the \lya transition,
even after allowing for the foreground \lya absorption at lower
redshifts.
In this subsection, we put a much stronger limit on $\Gamma_{-12}$ based on
the \lyb measurement of Paper II. 
We also attempt to use the \lyg trough to set a limit
on $\Gamma_{-12}$, although  it  suffers from larger uncertainties in the
modelling due to the presence of both foreground \lyb and \lya
absorption, and it is contaminated by \lyd absorption.

The \lyb absorption at redshift $z_{\beta}$ overlaps with the foreground
\lya absorption at the redshift
\begin{equation}
z_{\alpha} = (\lambda_\alpha / \lambda_\beta) (1+z_{\beta}) - 1.
\end{equation}
where $\lambda_{\alpha,\beta}$ are the rest-frame wavelengths of the 
Ly$\alpha$ and Ly$\beta$ transitions, respectively.
The observed total optical depth at the corresponding wavelength is
$\tau_{\rm total} = \tau_\alpha(z_\alpha) + \tau_\beta(z_\beta)$.
For a fractional density $\Delta$, 
\begin{equation}
\tau_{\beta} (\Delta) = (f_\beta / f_\alpha) \tau_{\alpha} (\Delta) = \tau_{\alpha}(\Delta) / 5.27,
\end{equation}
where $f_\alpha$ and $f_\beta$ are the oscillator strengths of \lya
and \lyb transitions, respectively, and $\tau_\alpha(\Delta)$ can be calculated using Eq.(11).
Therefore, the observed transmitted flux ratio in the presence of both
\lya and \lyb absorption can be written as,
\begin{equation}
 \begin{array}{lll}
	{\cal T}_{\alpha+\beta} & = \langle e^{-\tau_{\rm total}} \rangle & =
	\int_{\Delta_\alpha} \int_{\Delta_\beta}
	e^{-[\tau_\alpha(z_\alpha, \Delta_\alpha) + \tau_\beta(z_\beta, \Delta_\beta)} 
       p(z_\alpha, \Delta_\alpha) p(z_\beta, \Delta_\beta) 
	d\Delta_\alpha d\Delta_\beta\\
 & & = [\int_{\Delta_\alpha} e^{-\tau_\alpha(z_\alpha, \Delta_\alpha)}p(z_\alpha, \Delta_\alpha) d\Delta_\alpha]
       [\int_{\Delta_\beta} e^{-\tau_\beta(z_\beta)}p(z_\beta, \Delta_\beta) d\Delta_\beta]\\
 & & = {\cal T}_{\alpha}(z_\alpha) {\cal T}_{\beta}(z_\beta),
 \end{array}
\end{equation}
where $ {\cal T}_{\alpha}(z_\alpha)$ is the transmitted flux of the foreground
\lya absorption at $z_\alpha$ defined in Eq.(13), and
${\cal T}_{\beta}(z_\beta)$ is the transmitted flux due solely to the
\lyb absorption.

In Paper II, we found ${\cal T}_\beta = -0.002 \pm 0.020$ in
the \lyb GP trough of SDSS 1030+0524 at $z_{\rm abs} = 6$,  
after correcting for the foreground \lya absorption 
(${\cal T}_\alpha = 0.12$) at $z_{\rm abs} \sim 5.0$.
Using Eqs.(15) and (16) above, we find $\Gamma_{-12} < 0.025$ from 
the \lyb trough, an upper limit a factor of $\sim 3$ lower than that
estimated from the \lya trough.

The oscillator strength of the \lyg transition is a factor of 14.3
smaller than that of \lya and a factor of 2.7 lower than that 
of \lyb; therefore in principle, \lyg can be used to put an even 
stronger constraint on the equivalent Ly $\alpha$ optical depth.
In practice, the modelling of \lyg absorption is 
complicated due to three reasons:
\begin{enumerate}
\item Foreground \lya absorption at $z_{\alpha} \sim 4.6$.
We assume ${\cal T}_\alpha \sim 0.20$, following the values in Figure 1.
\item  Foreground \lyb absorption at $z_{\beta} \sim 5.7$.
From Figure 2, we find $\Gamma_{-12} (z=5.7)  \sim 0.12$, 
and the \lyb transmission ${\cal T}_\beta = 0.33$ using Equations (13) 
and (15).
\item Contamination from \lyd absorption.
\end{enumerate}
In the redshift range of the GP trough
$5.95 < z_{abs} < 6.16$, \lyg absorption covers a wavelength range
6759\AA\ $< \lambda < $ 6963\AA.
The \lyd absorption covers a wavelength range 6600\AA\ $< \lambda <$ 6800\AA,
partially overlapping with the \lyg trough, and 
the \lyd emission line is at $\lambda = 6914$\AA. 
From the Keck/ESI spectrum presented in Paper II, we find that
${\cal T} = 0.0043 \pm 0.0033$ in the wavelength range 
6759\AA\ $< \lambda < $ 6963\AA, and ${\cal T} = 0.0028 \pm 0.0041$
in the more restricted wavelength range 6800\AA $< \lambda < $ 6963\AA.
The latter wavelength range is redward of the \lyd GP trough. 
Although part of it is still bluer than the \lyd emission,
the amount of  \lyd absorption is strongly suppressed due to the 
proximity effect from the quasar.
In the wavelength range of 6914\AA\ $< \lambda < $ 6963\AA, 
totally redward of \lyd emission, we find
${\cal T} = -0.0022 \pm 0.0087$. 

From a combination of these transmitted flux ratio measurements, we adopt
the limit ${\cal T}_{\alpha + \beta +\gamma} < 0.007$ at 1-$\sigma$ in the
Ly$\gamma$ GP trough region.
Correcting for the foreground \lya and \lyb absorptions, 
we find the transmission solely due to \lyg to be
${\cal T}_\gamma < 0.10$.
This transmitted flux ratio corresponds to a photoionization rate
of $\Gamma_{-12} < 0.020$, comparable to the limit derived from the
\lyb GP trough, and representing a factor of $>20$ decline from
the ionizing background at $z\sim 3$.
The rapid decrease of $\Gamma$ at $z>5.7$ indicates a possible sharp transition
in the ionization state of the IGM (\S5).

\section{Simulating the \lya forest}

In this section, we use an N-body simulation to 
simulate the absorption spectra of high-redshift quasars in the
\lya region, taking into account the resolution and noise properties 
of the observations.
We fix the photoionization rate in the simulation to reproduce the
observed mean transmitted flux at each redshift, and compare the statistical 
properties of the  simulated spectra including
the probability distribution function (PDF) and the threshold crossing
statistics (\cite{jordi96}), with those of the observed spectra.

We carry out the simulations in a LCDM model, with $\Omega_{m} = 0.3$, 
$\Omega_{\Lambda} = 0.7$, $h=0.65$,
$\Omega_{b}h^{2} = 0.02$, and $\sigma_{8m} = 0.9$. Here, 
$\sigma_{8m}$ is the rms density fluctuation in $8 \hmpc$ spheres, chosen 
here to reproduce the
observed abundance of clusters at $z=0$ (\cite{white93}; \cite{eke96}).
We evolve the density fluctuations using a particle-mesh (PM) N-body 
code that is described in detail in Hennawi et al. (2001).
This code uses a staggered mesh to compute forces on particles 
(\cite{melott86}; \cite{park90}), and 
uses the leapfrog scheme described in Quinn et al. (1997)
to integrate the equations of motion.
The periodic simulation cube is $L=25 h^{-1}$Mpc on a side, and uses
$N_p=256^{3}$ particles and an $N_m=512^{3}$ force mesh.
We set up initial density fluctuations using the power spectrum
parameterization of Efstathiou, Bond \& White (1992) with the shape parameter
$\Gamma = 0.2$, and assign initial displacements and velocities to 
the particles using the Zeldovich approximation. We evolve from redshift 
$z=49 \rightarrow 5.45$, taking 55 equal steps in the expansion scale 
factor.


We use the TIPSY\footnote{see http://www-hpcc.astro.washington.edu/tools/tipsy/tipsy.html} package to 
create artificial \lya absorption spectra along 400 random lines of 
sight through the simulation cube.
TIPSY calculates the local mass density $(\rho_{m})$ at the location of each
dark matter particle using a cubic spline smoothing kernel
(Hernquist \& Katz 1989) enclosing 32 neighbors, and assigns it a
temperature $T = T_0\Delta^{0.6}$.
We assumed $T_0 = 10^{4}\,$K while creating the spectra, although the
values of $\Gamma_{12}$ we quote in this paper have been rescaled to
$T_0 = 2 \times 10^{4}\,$K.
Spectra along various lines of sight are then created from this particle 
distribution using the algorithm described by Hernquist et al.\ (1996).
We smooth the simulated spectra with a Gaussian filter of 
smoothing length $\sigma_{v} = 28 \kms$ (corresponding to a Full Width
at Half Maximum of $66 \kms$), and bin them in pixels 
of width $35 \kms$, approximately twice the resolution of the Keck 
spectra. We then add noise to the simulated \lya forest spectra.
The noise in the Keck spectra is dominated
by the background sky and is a strong function of wavelength
around $8500$\AA. In order to correctly simulate this structure
in the noise, we replace the fluxes $(F)$ in each simulated spectrum 
by the values $F \rightarrow F + \sigma_{\rm obs}G(1)$,
where $\sigma_{\rm obs}$ is the noise array in contiguous pixels
starting from a random location in the Keck spectrum binned 
to about $35 \kms$, and $G(1)$ is a Gaussian random deviate with 
zero mean and unit variance. 
We scale all the optical depths at each redshift so that the mean 
transmitted flux ratio computed from all the 400 simulated spectra
match the observed values. 


We compute the \lya forest spectra at $z_{\rm abs} = 5.5, 5.7$ and
6.0, and compare them with the observed Keck spectra
of SDSS 1044--0130, SDSS 1306+0356 and SDSS 1030+0524 at the
same wavelength ranges.
Figure 3 shows the \lya forest at $5.4 < z_{\rm abs} < 5.6 $ and 
$5.9 < z_{\rm abs} < 6.1 $ from the
Keck spectra, and the artificial \lya absorption spectra 
along eight random lines of sight through the simulation.
Since the simulation cube is only $25 \hmpc$ on a side,
each simulated spectrum is of length $3506\ \kms$,
$3558 \kms$ and $3635\ \kms$ at $z_{\rm abs}=5.5, 5.7$ and $6$,
respectively.
The simulated spectra are qualitatively similar to the \lya forest region in 
the Keck spectra. 
Although they have the same mean transmitted flux by construction, the 
fluctuations in the simulated spectra arise from fluctuations in the 
mass density field in a LCDM model (except at $z_{\rm abs} = 6$ where
both the observed and the simulated spectra are consistent with noise).
The fact that the fluctuations
in the simulated spectra are qualitatively similar to those in the
Keck spectra is an indication that our physical description of the
\lya forest as arising from photoionized gas in a low-density IGM in a 
LCDM model is reasonable.

Figure 4 shows the probability distribution function (PDF) of the transmitted
flux ratio (\cite{jenkins91}; \cite{jordi96}; \cite{rauch97}), computed using 
the 400 artificial spectra.
The solid points in the three panels show the flux PDF measured from the 
Keck spectra of SDSS 1044--0125, SDSS 1306+0356 and SDSS 1030+0524
in a velocity range of 8500 km s$^{-1}$ around
 $\langle z_{\rm abs} \rangle =5.5$, $5.7$ and $6.0$, respectively.
The solid line shows the average flux PDF from the simulated spectra.
Since the simulated spectra are shorter in length than the data,
we compute the cosmic variance in the flux PDF measured from a single Keck
spectrum of length $8500 \kms$ by combining together $2.5$ artificial 
spectra end to end
so that the resulting artificial spectrum has roughly the same length
as the \lya forest region in the Keck spectrum.
The flux PDF of the simulated spectra is consistent with that computed
from the Keck spectra, in all three redshift bins.
In order to match the observed mean transmitted flux ratio, we find that 
the ionizing background at $z\sim 5.5$ and $5.7$ is
$\Gamma_{-12} = 0.21$, a factor of 2.5 lower than the value at $z\sim 3$,
and consistent with the result from \S3.1.
Note that the shape of this distribution is related to the shape of the
power spectrum, and is a prediction from the LCDM model.
For $z\sim 6.0$, the simulated spectrum reproduces the observed
GP trough with $\Gamma_{-12} < 0.08$, again consistent with
the limit derived in \S3.1 using a semi-analytic model.
At this redshift, since there is no detected flux in the Keck spectrum, 
the observed
and simulated PDFs merely reflect the distribution of noise in the spectrum.

The three panels in Figure 5 show the number of flux threshold crossings 
per unit redshift (the threshold crossing statistic; \cite{jordi96}) 
in the simulated spectra (solid line) and the Keck spectra (points) in three
redshift bins. We show this quantity as a function of the
fraction of pixels in the spectrum that are less than the threshold
flux (\cite{jordi96}; \cite{weinberg98a}).
This statistic is analogous to the ``genus curve''
used to characterize the topology of the 3-dimensional galaxy distribution
and has the advantage that it is insensitive to the exact relation between 
the relative distribution of dark matter and baryons (the ``biasing'' 
of the \lya forest), as long as this relation is monotonic.
We compute the cosmic variance in this statistic in a similar manner
to the flux PDF. 
The simulated spectra reproduce this statistic of the observations
very well, at all three redshifts.
Therefore this modeling of the \lya forest
in a LCDM model can reproduce the topology of the 
dark matter density fluctuations that are probed by 
the observed \lya forest in these quasar spectra.

\section{Constraining the Epoch of Reionization}

In this section, we expand the model described in \S3.1 to 
calculate from observations the evolution of three quantities that reflect
the ionization state of the IGM:
(1) the neutral fraction of the IGM, taking into account the inhomogeneity 
of the IGM,
(2) the critical density -- the density lower than which the gas is 
completely ionized,
and (3) the mean free path of ionizing photons in the IGM.
Using these calculations, we  show that as
reionization progresses toward lower redshifts, the neutral fraction 
decreases, regions of progressively higher densities become ionized, and 
the mean free path of ionizing photons increases.
We compare these three quantities with the results from a simulation of the 
cosmological reionization process by Gnedin (2000),  
to constrain the likely epoch of reionization (see also \cite{Gnedin01}).

We can use Eq.(6) to calculate the local neutral hydrogen fraction $[f_{\rm HI}(\Delta)]$ 
in a region of density $\Delta$,  
\begin{equation}
f_{\rm HI}(\Delta) = \frac{n_{\rm HI}(\Delta)}{n_{\rm HI}(\Delta) + 
n_{\rm HII}(\Delta)} = \frac{n_{\rm HI}(\Delta)}{\langle n_{\rm H} \rangle \Delta},
\end{equation} 
where $\langle n_{\rm H} \rangle$ is the average $n_{\rm H}$ over the universe.
We define
\begin{equation}
A = 1.16 \langle n_{\rm H} \rangle \alpha(T) / \Gamma,
\end{equation}
where the coefficient 1.16 comes from the helium contribution to
the electron density assuming a helium fraction $Y=0.24$ (Weinberg et al. 1997).  
Then, it can be shown that
\begin{equation}
f_{\rm HI}(\Delta) = \frac{-1+\sqrt{1+4A\Delta}}{1+\sqrt{1+4A\Delta}}.
\end{equation}
Note that for a highly ionized medium, $f_{\rm HI} \sim A\Delta$.
We calculate the redshift evolution of the HI fraction using
the density distribution of Eq.(13), and the results on $\Gamma$ in 
Figure 2.
Note that $f_{\rm HI}$ is directly related to $A \sim \alpha(T)/\Gamma$.
Therefore, our estimate of  $f_{\rm HI}$ is insensitive to our
assumed value for the IGM temperature $T$,
since our estimate of $\Gamma$ is itself proportional to the assumed $\alpha(T)$.
Figure 6 shows both the volume-averaged neutral hydrogen fraction 
 $f_{\rm HI}^{\rm V}$ (dashed 
line) and the mass-averaged quantity $f_{\rm HI}^{\rm M}$ 
(solid line).
The volume-averaged HI fraction increases from $1.7 \times 10^{-5}$ at $z\sim 3.9$ to $1.3\times10^{-4}$ at $z=5.8$ -- an increase 
by a factor of $\sim 8$.
In the GP trough region at $z\sim 6.05$, the upper limit on the 
ionization rate $\Gamma$ from the \lya, \lyb and \lyg 
transitions yields a {\em lower limit} on the neutral fraction,
$f_{\rm HI}^{\rm V} \gtrsim 10^{-3}$, and 
$f_{\rm HI}^{\rm M} \gtrsim 10^{-2}$, an increase by almost two orders
of magnitude from $z\sim 4$.
The mass-averaged value is larger because most HI is concentrated in 
dense clumps in the IGM which occupy a small volume.

The neutral hydrogen fraction depends on 
the detailed density distribution of the IGM.
As pointed out in \S3.1, $\tau_{\rm eff} < \langle \tau \rangle$.
Therefore, simply using Eq.(2), which assumes  a homogeneous IGM,
will severely underestimate the neutral fraction. 
For example, at $z=5.8$, $\tau_{\rm GP}^{\rm eff} = 3.7$ (Figure 1), 
and the neutral fraction derived using Eq.(2) is $5.6\times 10^{-5}$;
a factor of $\sim 2.3$ and $\sim 70$ lower than the volume and mass-averaged
neutral fractions, respectively, from Eq.(19). 
In a clumpy IGM, the high-density regions quickly become
opaque to all \lya photons, while almost all the transmitted flux comes 
through the deepest voids in the IGM. 
We illustrate this point in Figure 7, 
which shows the cumulative distributions of transmitted flux as a
function of density at z=5.3.  At this redshift, while $\tau_{\rm eff}
\sim 2.5$, the region with the minimum absorption in the spectrum of
SDSS 1044--0125 (Fan et al. 2000) has $\tau < 0.5$. It represents the
deepest void in the IGM, with $\Delta \sim 0.15$ (from Eq. 11), and has
a neutral fraction of $\sim 1.2 \times 10^{-5}$, a factor of seven lower
than the volume-averaged neutral fraction at this redshift. It is a factor
of $\sim 50$ lower than the value one would get by directly applying
Eq.(2) and assuming a uniform IGM, as $\tau$ scales as $\Delta^2$ (Eq. 11).

In all the calculations above, we have assumed that the ionizing background
(and hence $\Gamma$) is uniformly distributed in the IGM.
Figure 7 shows that the $\Gamma$ derived from the observed transmitted flux 
ratio is characteristic of the background level in low density regions of 
the IGM, since the high density regions contribute little to the observed flux.
The observed transmitted flux [Eq.(11)]
is essentially a volume-averaged measurement. In a similar sense, our 
calculation of 
the volume-averaged neutral fraction should be regarded as more
reliable than the mass-average, as the former quantity is less affected by the 
the local ionization state in the high density regions.

Miralda-Escud\'e et al. (2000) present a semi-analytic model describing the 
evolution of 
reionization in an inhomogeneous universe; reionization starts in voids 
and then gradually penetrates deeper into overdense regions.
This process is defined by the redshift evolution of the critical density
parameter $\Delta_i$.
At any redshift, regions with $\Delta > \Delta_i$ remain neutral,
while regions with $\Delta < \Delta_i$ are mostly ionized.
Here we 
follow Gnedin (2000) by defining $\Delta_i$ 
as the critical density above which 95\% of the neutral gas lies.
Figure 8 shows the evolution of $\Delta_i$ based on the neutral fraction
calculated above. The quantity $\Delta_i \sim 40$ at $z\sim 4$, and it quickly 
decreases to $\Delta_i < 4$ in the GP trough region at
$z \sim 6$.
Using Eq.(8) of Miralda-Escud\'e et al. (2000),
we calculate the mean free path of the ionizing photons, 
$\lambda_i = \lambda_0 [1-F_{\rm v}(\Delta_i)]^{-2/3}$,
where  $F_{\rm v}(\Delta_i)$ is the fraction of volume with 
$\Delta < \Delta_i$, and 
$\lambda_0 H$ = 60 km s$^{-1}$  (Miralda-Escud\'e et al.  2000). 
Figure 9 shows the evolution of the comoving mean free path with redshift.
It decreases from $\sim 80 \hmpc$ at $z\sim 4$ to 
smaller than $\sim 8 \hmpc$ in the GP trough region at $z \sim 6$.
 
Figures  6, 8 and 9 clearly demonstrate that at $z\sim6$, the universe is much more
neutral than at lower redshift ($z\sim 3 - 5$). 
At $z\sim 6$, the presence of the GP trough shows that more than 1\% of
the protons are in the form of HI,
the moderately overdense regions ($\Delta \gtrsim 4$) are
still mostly neutral, and the mean free path of ionizing photons is several 
Mpc in comoving distance --- a distance comparable to the 
correlation length of the galaxy distribution at $z = 0$.
The last remaining HI in the IGM 
is being ionized at this epoch.
In this sense, the first appearance of the complete GP trough at
$z\sim 6$, marks the {\it end of the reionization epoch}.

Cosmological reionization is a complicated process. Cosmological
simulations that include gas dynamics, star formation processes,
atomic and molecular physics and radiative transfer have recently begun to
to shed light on when and how this process happened in popular 
cosmological models.
Gnedin (2000) divides the reionization process into three phases: 
the pre-overlap stage in which individual HII regions begin to grow; the 
overlap stage in which these HII regions merge and the ionizing 
background in the IGM
rises by a large factor; and the post-overlap stage in which 
the remaining HI in
the high density regions is gradually ionized.
The results from the simulation of Gnedin (2000)
are shown in Figures 6, 8 and 9.
In the simulation,  the overlap stage was assumed to be at $z\sim 7$.
At this redshift, the universe goes through a near phase-transition: 
the ionizing background and the mean free path of photons increases and the neutral hydrogen
fraction decreases dramatically over a narrow redshift range.
The reionization redshift can be defined in a number of ways: 
e.g., the redshift at which the average neutral hydrogen fraction is 
$\sim 50$\%; the redshift at which the HII filling factor is of order 
unity;  the redshift at which
an average region of the IGM is ionized
($\Delta_i \sim 1$), etc.
However, as shown in Gnedin (2000), Chiu \& Ostriker (2000) and
Razoumov et al. (2001), the overlapping stage occurs over a small range in
redshift, with the neutral fraction changing from 90\% to $<1$\% in 
$\Delta z< 2$
(corresponding to a time interval of 200 million years at $z \sim 7$),
while $\Gamma_{-12}$ increases from $\sim 10^{-3}$ to $\sim 10^{-1}$.

Comparing the predictions from the simulation with the observations
in Figures 2, 6, 8 and 9, it is evident that the redshift range $z<5.7$ is 
consistent with being in the post-overlap stage of reionization, where 
$\Delta_i$ increases as the remaining overdense regions in the IGM are 
being ionized.
The emergence of the GP trough at $z\sim 6$, and the stringent lower limits on the neutral 
hydrogen fraction from the presence of the \lyb and \lyg GP troughs
suggest the onset of a rapid transition in the ionization state of the IGM.
The upper limits on the ionizing background and mean free path, and
the lower limits on the neutral fraction and critical overdensity we
derive from the observations are all consistent with the typical
values near the end of the overlap stage of reionization 
in the simulations.
In this sense, the epoch of $z\sim 6$ is at the transition from the 
overlap stage to the post-overlap stage of reionization. 
Even though the current observations certainly cannot probe deeper into
the reionization epoch, the near phase-transition behavior of the 
ionization state of the IGM at $z\sim 6$, and the narrow redshift range
over which this process occurs in cosmological reionization simulations
both imply that the reionization redshift
{\em cannot be at a redshift much higher than six}. 

We caution, however, that these results are based on the GP trough in
{\em only one} quasar, SDSS 1030+0524.  The IGM is unlikely to be reionized
in an uniform manner.  We have shown that at $z\sim 6$, the mean free path
of the ionizing photons is likely to be shorter than 10 comoving Mpc.
In this case, the ionizing background could show substantial fluctuations.
These fluctuations could be due to the rarity and the possible strong
clustering of ionizing sources, as well as the radiative transfer effects
such as shadowing of the sources. For example, in the simulation of
Gnedin (2000), the ionizing background shows fluctuation of a factor of
a few at $z\sim 6$ (his Figure 5) in the underdense regions of the IGM
(which occupy most of the volume). In order to calculate the amplitude of
these fluctuations, we would need detailed simulations of the formation
of the first generation of ionizing objects. This is beyond the scope of
the present paper The first generation of objects is likely to be highly
biased and clustered, which would give rise to different reionization epochs
along different lines of sight and in different regions of the IGM.
If the universe was ionized by luminous quasars (unlikely based on
current statistical on quasar evolution, Fan et al. 2001c), the large HII
regions produced by them near the epoch of overlap could even lead to gaps
in transmitted flux (Haiman \& Loeb 1998, Miralda-Escud\'e et al. 2000).
Thus, it is quite possible that the next quasar discovered at $z>6$ will
show somewhat different absorption properties.

\section{Discussion}


The presence of the GP trough in the spectrum of the highest redshift
quasar provides a first opportunity to study 
structure formation at high-redshift and 
the formation of the first generation of stars and galaxies.
To first order, the redshift of the reionization epoch depends on 
two factors:
(a) the amount of small scale power in the power spectrum of mass density
fluctuations, which determines how many halos can collapse to
make stars at a certain redshift.
It also determines the clumpiness of the IGM;
and (b) how efficiently the collapsed halos can produce UV photons,
which in turn depends on the stellar initial mass function, 
the efficiency of star formation, the escape factor of UV photons from
proto-galaxies, etc.
The reionization redshift and the observed ionizing background
can be used to put constraints on these factors.

In \cite{CO00}'s semi-analytic calculations,
the reionization redshift for low density $\Lambda$-dominated CDM
models is typically in the range of 7 -- 11, not very far from
$z \sim 6 - 8$ suggested in this paper.
On the other hand, most of the non-Gaussian texture and isocurvature models
have too much small scale power and a much earlier reionization redshift
$z \gtrsim 10$, making them difficult to reconcile with the observations
presented in this paper.
Barkana, Haiman \& Ostriker (2001) used the constraint on the
reionization epoch to put limits on models of Warm Dark Matter.
Assuming that the reionization redshift is smaller than 7, we find
that the mass of the warm dark matter particle must be smaller than
3 keV in the standard model they considered.

Paper I estimates that the SDSS will find about 20 luminous quasars 
in the redshift range $6 < z < 6.6$ over the 10,000 deg$^2$ survey area.
High resolution, high signal-to-noise ratio spectra of these
luminous quasars in the Lyman series absorption regions
will provide valuable probes of the end of the reionization epoch, 
and in particular measure the spatial inhomogeneity of the reionization 
process.

In order to probe deep into and beyond the reionization epoch,
souces at higher redshift are needed.
The \lya emission line moves out of the optical window and into the
infrared at $z\sim 6.6$. Thus, the objects become very faint in the
optical wavelengths due to Lyman series absorption by  neutral hydrogen
gas in the foreground at lower redshifts.
Hence, this is the limiting redshift for discovering sources
from ground-based optical surveys.
Deep and wide-field IR surveys such as the
PRIME\footnote{http://prime.pha.jhu.edu/index.html} mission are required
to find objects at even higher redshifts.

The stringent limits on the neutral hydrogen fraction at
low redshifts from the GP test arise from the large cross section
of neutral hydrogen to \lya photons. For the same reason, the
GP test quickly becomes insensitive to larger neutral hydrogen fractions.
With a half hour exposure on the Keck telescope, we are able to place
a lower limit on the neutral fraction $f_{\rm HI} \gtrsim 10^{-2}$.
As the neutral fraction hydrogen scales with $\tau$, or logarithmically
with the flux, it is impossible to obtain a limit more than a few times
stronger than the current limit for any reasonable exposure time.
Therefore, the original version of the GP test cannot be used to probe
deeper into the reionization epoch, for which $f_{\rm HI} \sim 1$.

For an object observed prior to the reionization epoch,
the damping wing of the GP trough arising from the large GP
optical depth of the neutral medium will extend  into the
{\em red} side of the \lya emission line (Miralda-Escud\'e 1998).
The depth and extent of this damping wing can in principle  be used
to measure the neutral fraction and the redshift of the reionization epoch.
However, this GP damping wing test cannot be applied
to luminous quasars, due to the proximity effect from the quasar itself.
As shown by Madau \& Rees (2000) and Cen \& Haiman (2000), the quasar will
ionize a HII region
around it if the bulk of the IGM is still neutral, resulting in a
line profile very similar to the case where the IGM has already been ionized.
An alternative is to search for lower luminosity quasars (Stern et al. 2000) 
and galaxies (\cite{Hu}, \cite{RM01}) at high-redshifts by either applying 
the color  dropout technique to deep multi-band imaging data or by search
in  regions around foreground galaxy clusters where background objects 
might be amplified by gravitational lensing (e.g \cite{Ellis01}).
The highly magnified sources may be bright enough to allow high S/N
spectroscopy, necessary for accurately measuring the profile of the
damping wing.

Another promising opportunity for probing the reionization epoch arises
from the observations of IGM absorption in the afterglow of 
high redshift Gamma Ray Bursts (GRBs, Loeb 2001). 
GRBs are transient phenomena with no immediate impact on the surrounding 
IGM. The time dilation at high-redshift leads to a longer duration for
the afterglows, making the identification and follow-up 
observations easier.
Moreover, if high-redshift GRBs are associated with the collapse of 
massive stars, their evolution is likely to be similar to those of
star-forming galaxies, which show slower decline at high redshift
(\cite{Steidel98}) than the rapid decline in the number density
of quasars (e.g. Fan et al. 2001b).

We thank Renyue Cen, Zoltan Haiman, Avi Loeb, Pat McDonald, Jerry Ostriker, 
Martin Rees, Joop Schaye, and David Weinberg for helpful discussions.
We acknowledge support from NSF grant PHY00-70928 and
a  Frank and Peggy Taplin Fellowship (XF), and NSF grant
AST-0071091 (MAS).

\newpage
\begin{figure}
\vspace{-2.2cm}

\epsfysize=600pt \epsfbox{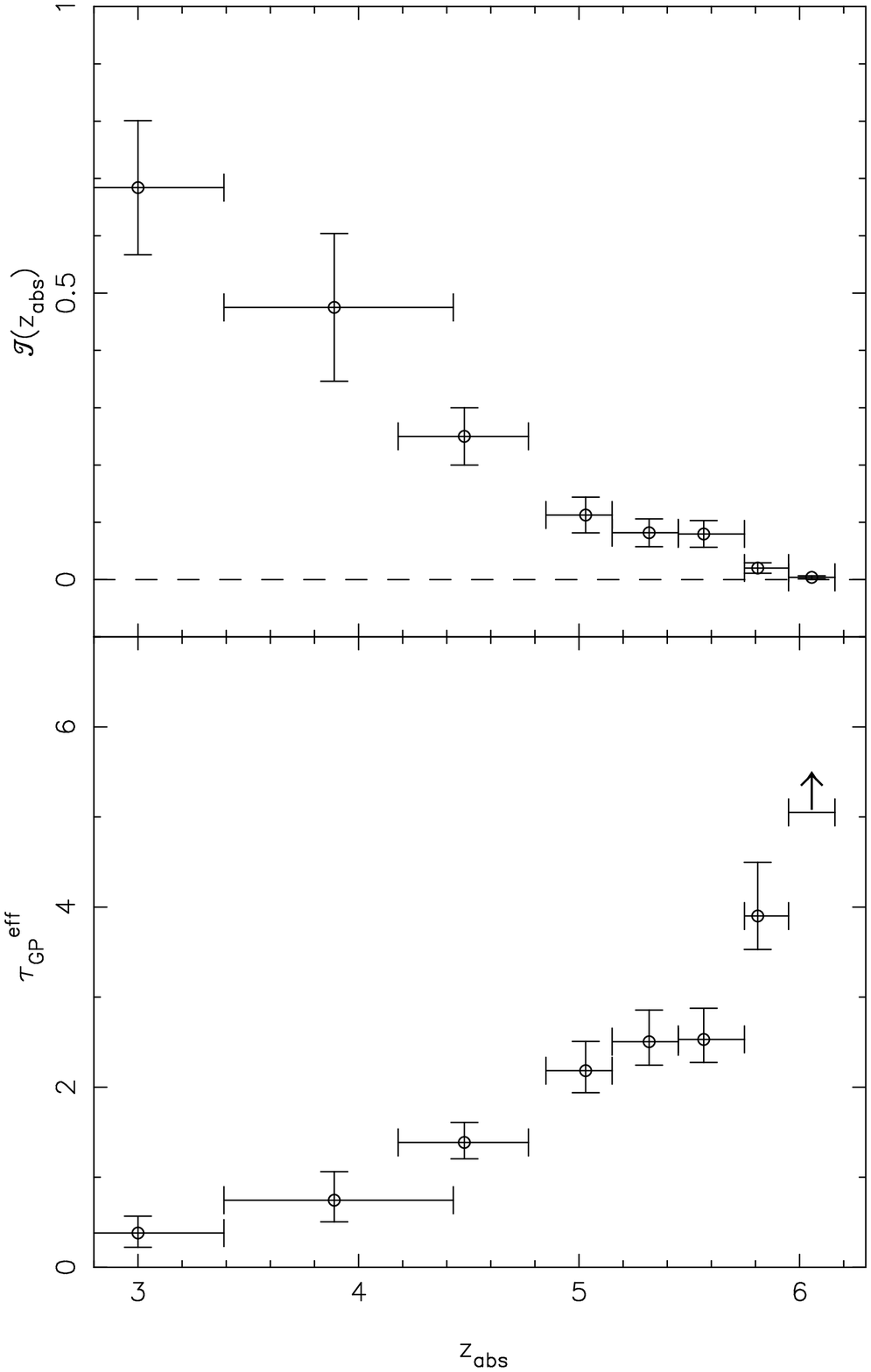}

Figure 1. Evolution of \lya absorption
based on the observations of four quasars at $z>5.7$ in
Fan et al. (2001), Becker et al. (2001) and Pentericci et al. (2001).
The results at $z_{\rm abs} < 5.6$ are averaged over four lines of sight,
 and the error bars include contributions from photon noise, 
uncertainty in the intrinsic quasar continuum, and 
an estimate of the sample variance.

\end{figure}

\begin{figure}
\vspace{-2cm}

\epsfysize=600pt \epsfbox{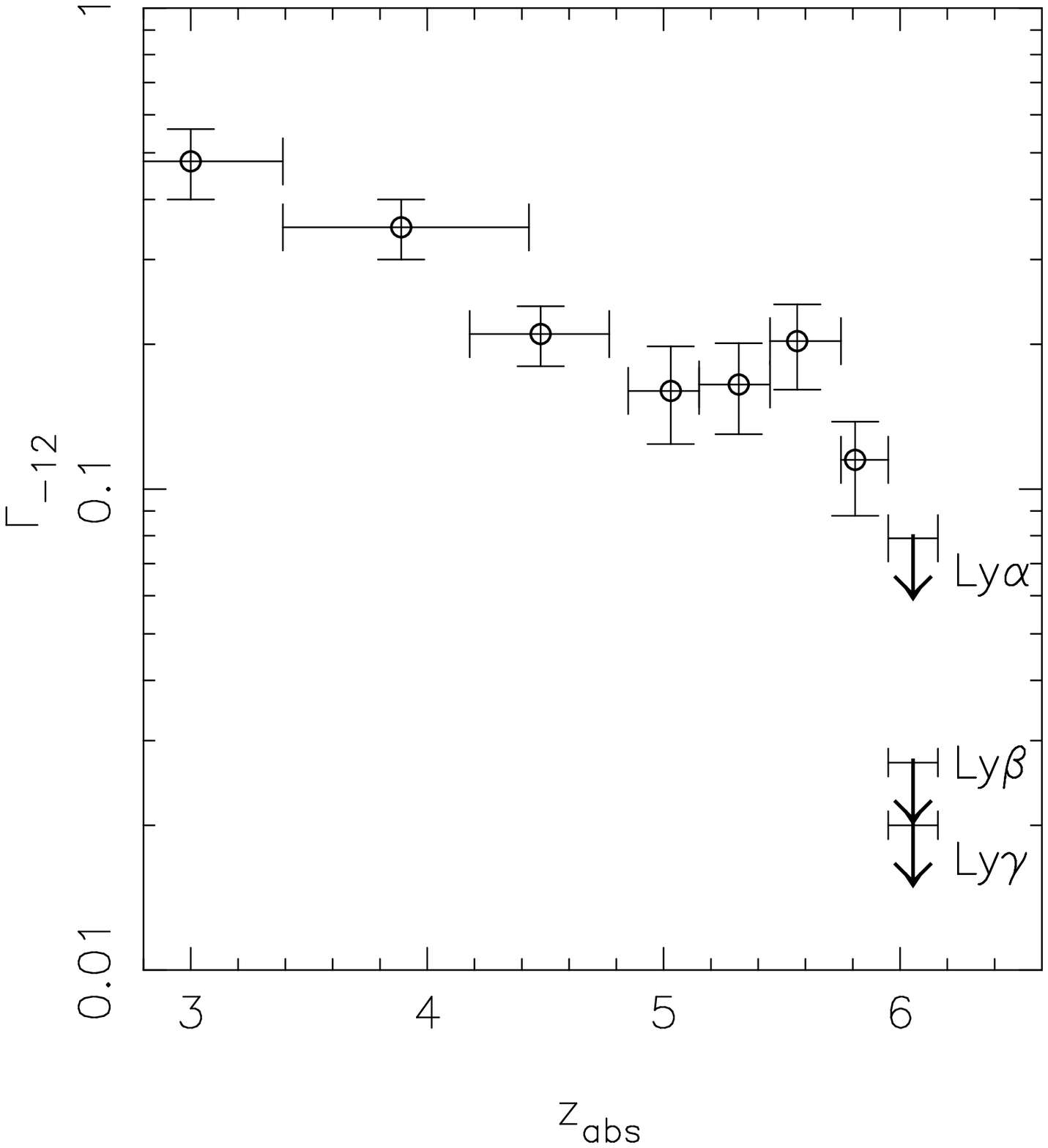}

Figure 2. Evolution of the photoionization rate (in units of
$10^{-12}\ {\rm s}^{-1}$) with redshift. 
The points at $z_{\rm abs} < 4.5$ are taken from \cite{MM01}. 
For the highest redshift measurement (at $z = 6.05$), the upper limits
based on \lya, \lyb and \lyg GP troughs are
shown separately.

\end{figure}

\begin{figure}
\centerline{
\epsfxsize=\hsize
\epsfbox[18 50 592 738]{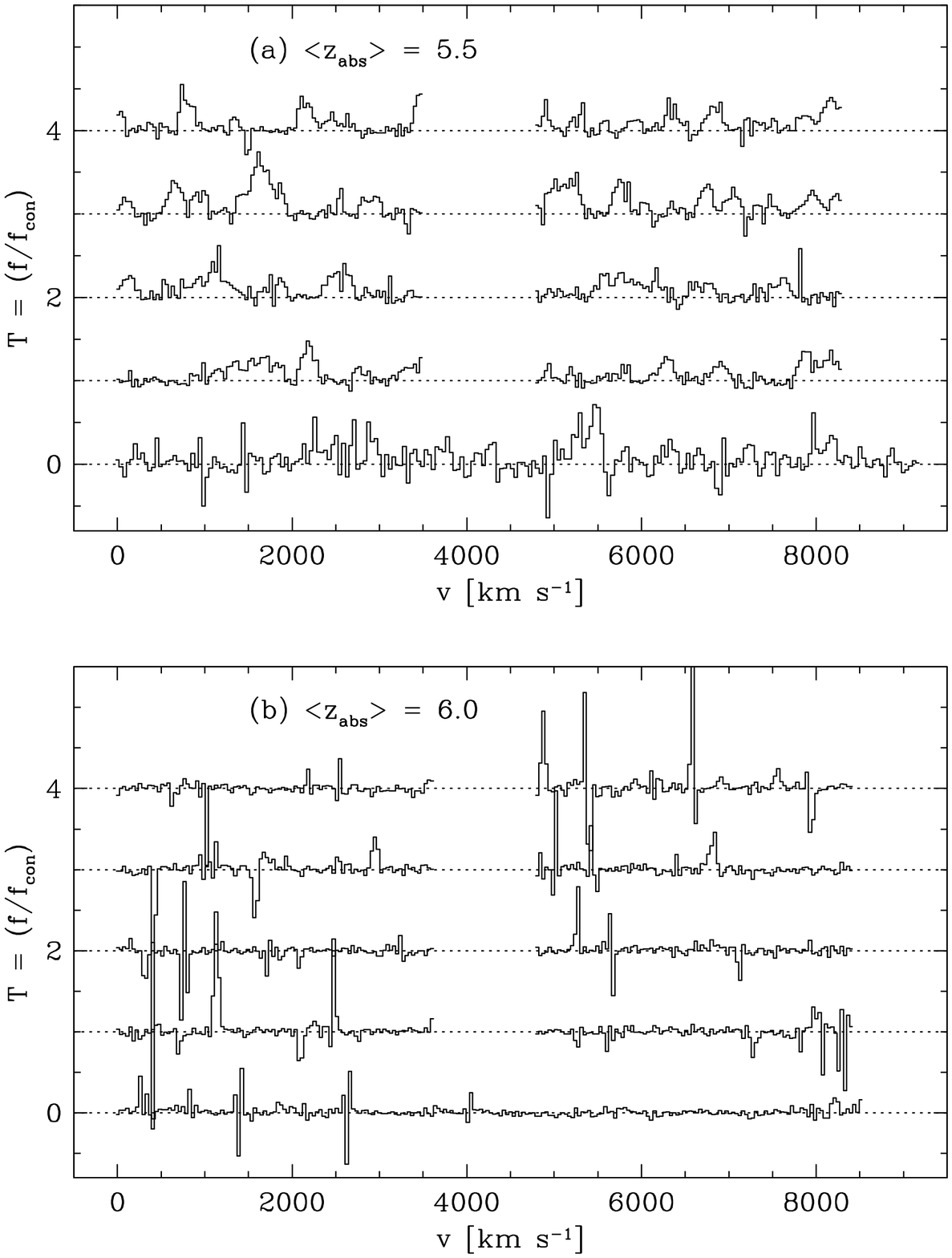}
}

Figure 3. Comparison of \lya forest spectra at
(a) $z_{\rm abs} = 5.5,$ and (b) $z_{\rm abs} = 6.0$. In each panel, 
the spectrum at the bottom shows the transmitted flux divided by a model
for the continuum, in the \lya forest region at that redshift in the
Keck spectrum of SDSS 1044--0125 and SDSS 1030+0524, respectively.
The smaller chunks are simulated spectra  along eight different lines of 
sight through the simulation cube, with the same resolution, binning and 
noise properties as the Keck spectra. Pairs of simulated spectra are offset 
by one unit in the vertical direction for clarity. 
\end{figure}

\begin{figure}
\centerline{
\epsfxsize=\hsize
\epsfbox[18 144 592 718]{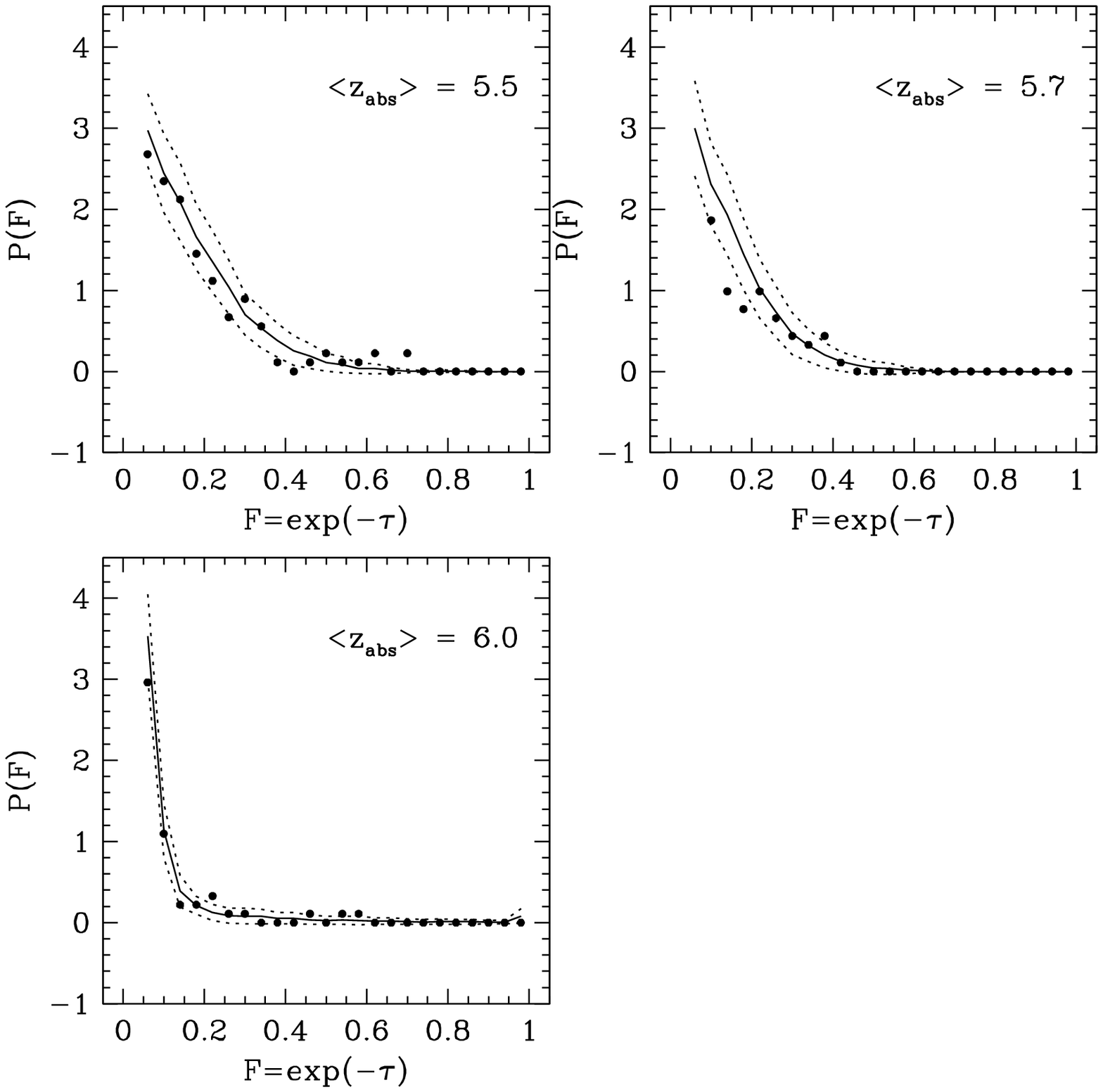}
}

Figure 4. Probability distribution function (PDF) of the transmitted
flux in the \lya forest region in the Keck spectra (points) and
simulated spectra (lines) at (a) $z_{\rm abs}=5.5,$ (b) $z_{\rm abs} = 5.7$
 and (c) $z_{\rm abs} = 6.0.$
The solid line shows the average flux PDF computed using 400 artificial
spectra, and the dashed lines shows the expected $1-\sigma$
uncertainty in the PDF from a single spectrum of length equal to that of
the real spectrum (i.e., the cosmic variance).
\end{figure}

\begin{figure}
\centerline{
\epsfxsize=\hsize
\epsfbox[18 144 592 718]{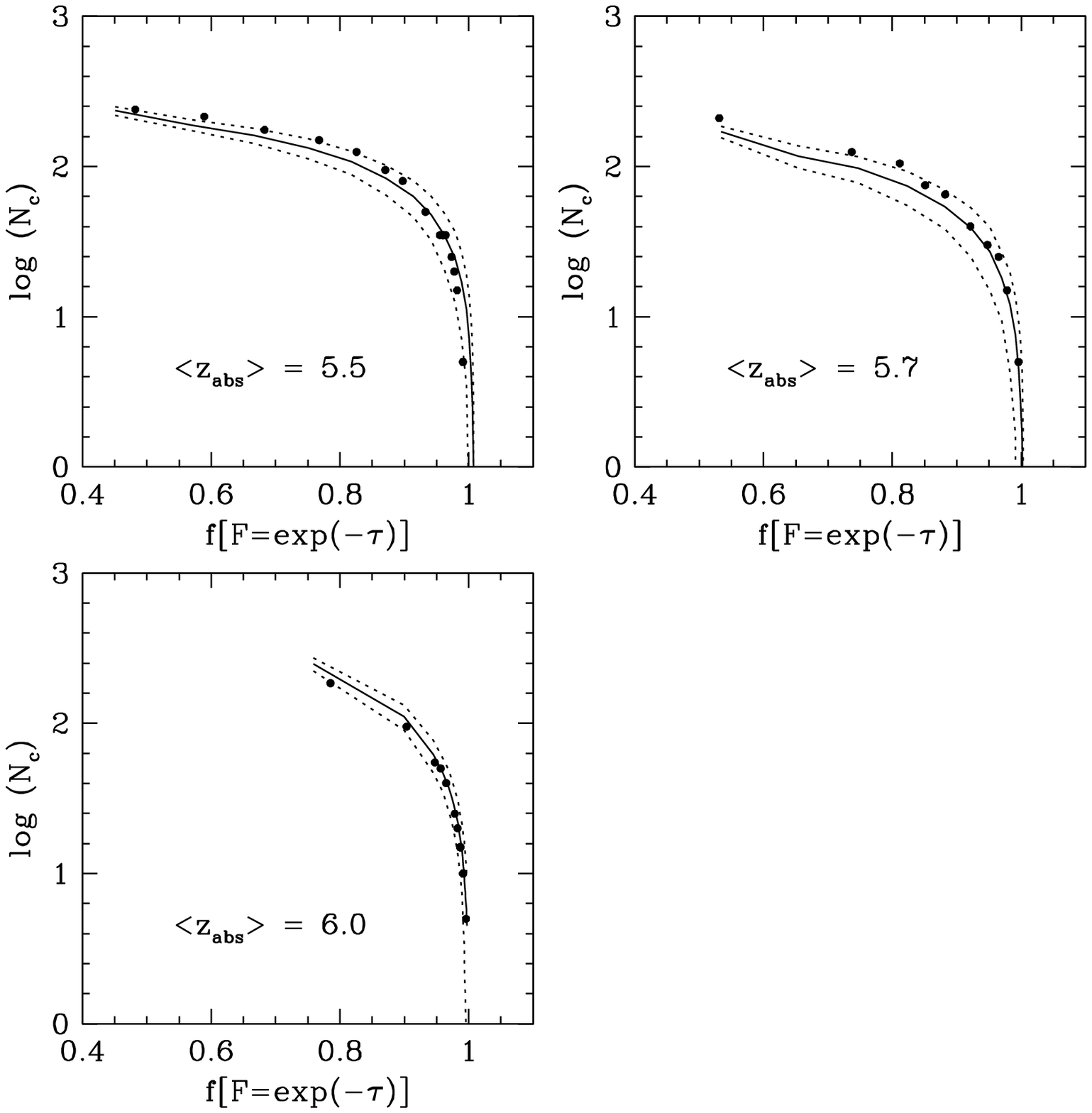}
}

Figure 5. 
Threshold crossing statistic as a function of the fraction of
pixels in the spectrum less than the flux threshold in
the \lya forest region in the Keck
spectra (points) and simulated spectra (lines) at (a) $z_{\rm abs}=5.5$,
(b) $z_{\rm abs} = 5.7$ and (c) $z_{\rm abs} = 6.0.$
The solid line shows this statistic computed using 400 artificial
spectra, and the dashed lines shows the expected $1-\sigma$
uncertainty in this quantity from a single spectrum of length equal to that of
the real spectrum (i.e., the cosmic variance).
\end{figure}

\begin{figure}
\vspace{-2cm}

\epsfysize=600pt \epsfbox{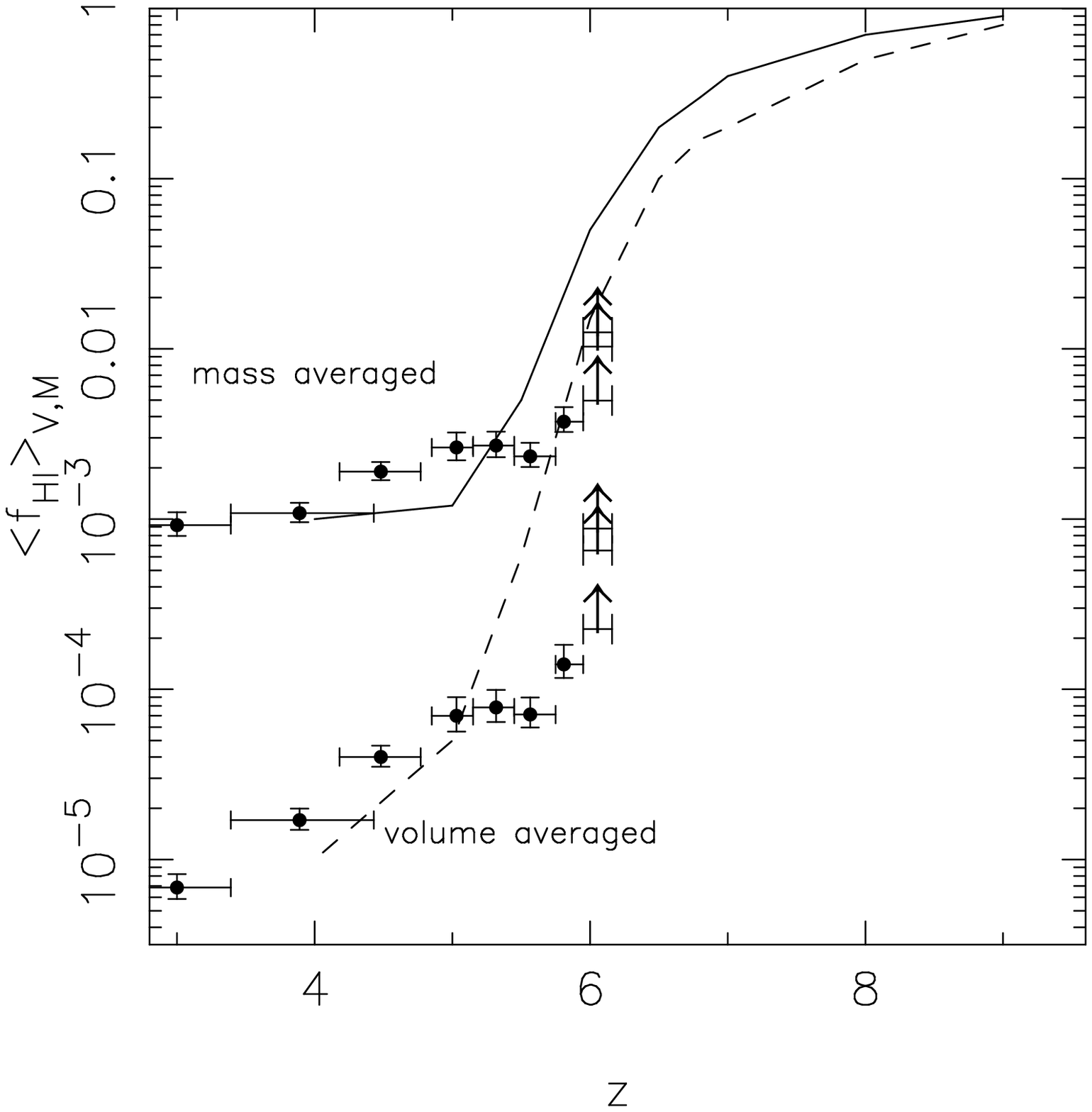}

Figure 6. Evolution of neutral hydrogen fraction of the IGM.
The solid points with error bars are measurements based on the
four high-redshift quasars, for both mass-averaged, and volume-averaged
neutral fraction. 
The solid and dashed lines are the mass- and volume-averaged
results from the N128\_L4\_A simulation of Gnedin (2000).
The neutral fraction inferred from the observations
is comparable to that of the
transition from overlapping stage to post-overlap stage of
reionization in the simulation.
\end{figure}

\begin{figure}
\vspace{-2cm}

\epsfysize=600pt \epsfbox{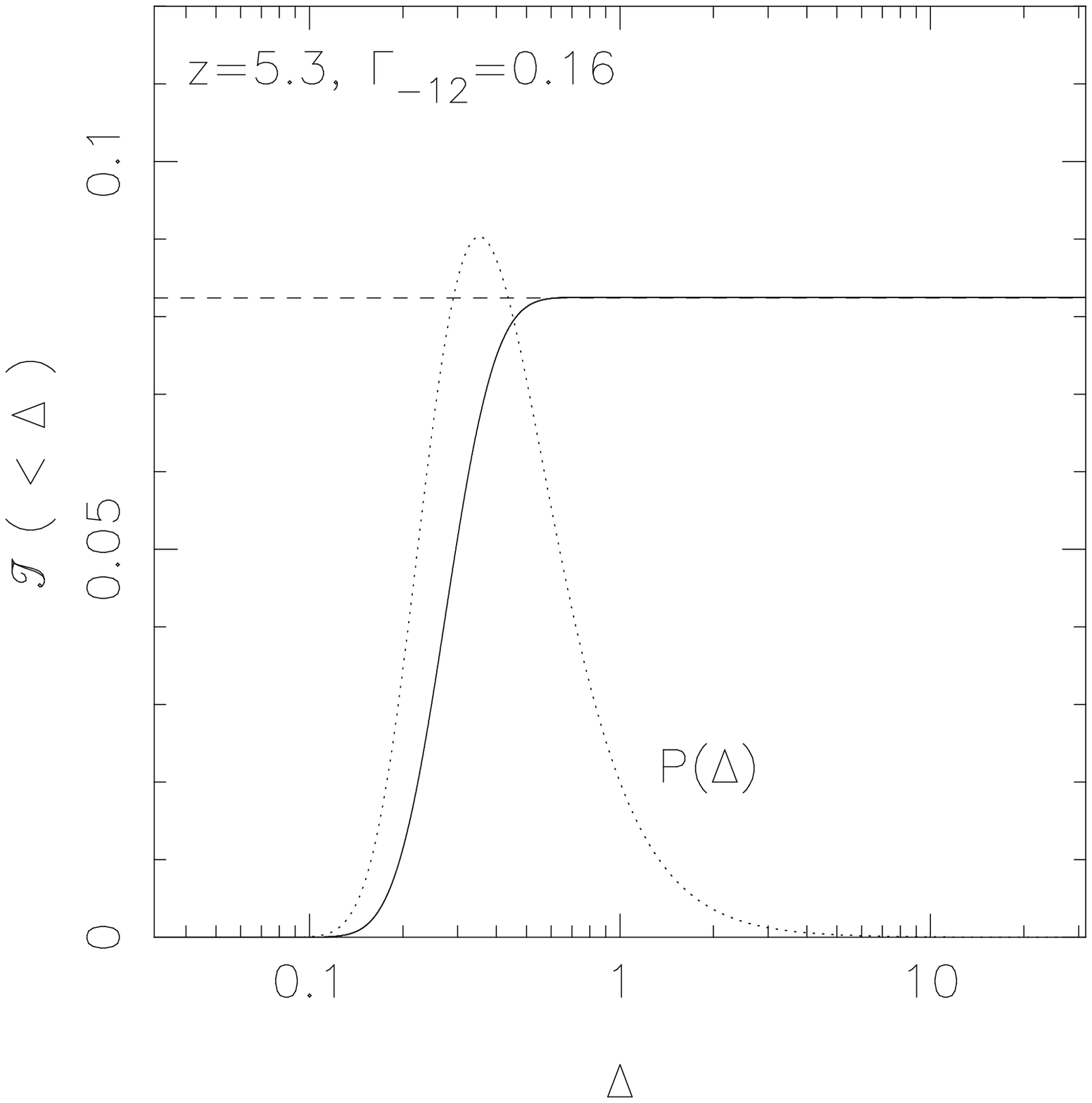}

Figure 7. Cumulative distribution of transmitted flux ratio as a function of 
density at $z=5.3$, with  a transmitted flux ratio of
0.08, corresponding to $\Gamma_{-12} = 0.15$. 
The dashed line shows the probability distribution function of
the density.
Most of the transmitted flux comes from the underdense
voids in the IGM.

\end{figure}

\begin{figure}
\vspace{-2cm}

\epsfysize=600pt \epsfbox{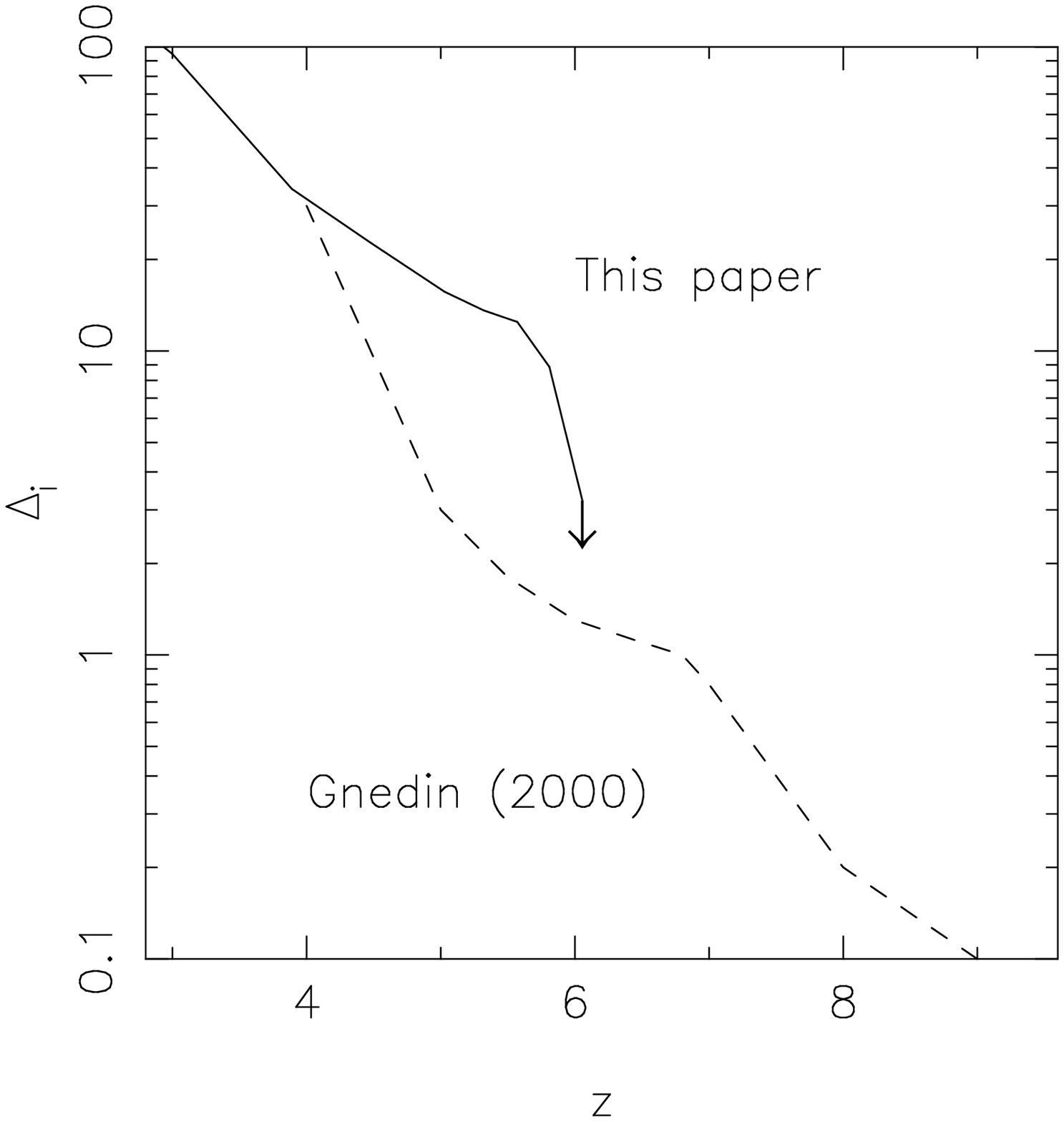}

Figure 8. Evolution of the limiting density, $\Delta_i$, defined as the 
density above which 95\% of the HI gas lies.
In this picture, reionization starts in voids, and gradually
penetrates deeper into overdense regions in an inhomogeneous IGM.
Both results inferred from the observations (solid line) and those from the simulation of
Gnedin (2000, dashed line) are shown.

\end{figure}

\begin{figure}
\vspace{-2cm}

\epsfysize=600pt \epsfbox{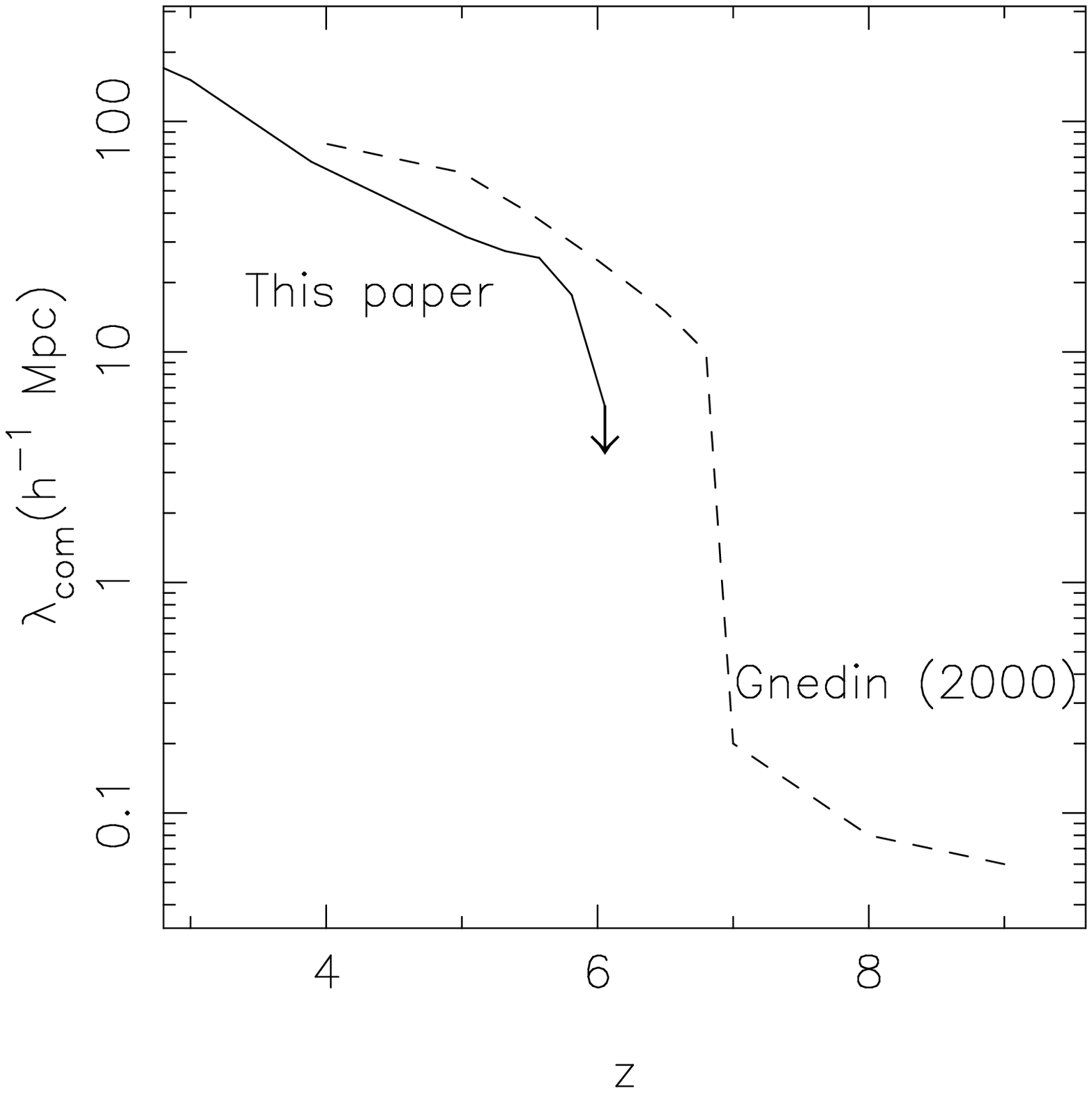}

Figure 9. Evolution of the mean free path of ionizing photons inferred
from the
observations (solid line) and simulation of Gnedin (2000, dashed line).
\end{figure}

\end{document}